\documentclass[pra,aps,preprint,showpacs]{revtex4}
\usepackage{epsfig}

\begin{document}

\title{Direct measurement of optical quasidistribution
functions: multimode theory and homodyne tests of Bell's
inequalities}

\date{\today}

\author{Konrad Banaszek}
\affiliation{Clarendon Laboratory, University of Oxford, Parks Road, Oxford OX1 3PU,
United Kingdom}

\author{Andrzej Dragan and Krzysztof W\'{o}dkiewicz}
\affiliation{Instytut Fizyki Teoretycznej, Uniwersytet Warszawski, Ho\.{z}a 69,
PL-00-681 Warszawa, Poland}

\author{Czes{\l}aw Radzewicz}
\affiliation{Instytut Fizyki Do\'{s}wiadczalnej, Uniwersytet Warszawski, Ho\.{z}a 69,
PL-00-681 Warszawa, Poland}

\begin{abstract}
We develop a multimode theory of direct homodyne measurements of quantum
optical quasidistribution functions. We demonstrate that unbalanced homodyning
with appropriately shaped auxiliary coherent fields allows one to sample
point-by-point different phase space representations of the electromagnetic
field. Our analysis includes practical factors that are likely to affect the
outcome of a realistic experiment, such as non-unit detection efficiency,
imperfect mode matching, and dark counts. We apply the developed theory to
discuss feasibility of observing a loophole-free
violation of Bell's inequalities by measuring
joint two-mode quasidistribution functions by photon counting
under locality conditions.
We determine the range of parameters of the experimental setup that
enable violation of Bell's inequalities for two states exhibiting entanglement
in the Fock basis: a one-photon Fock state divided by a 50:50 beam splitter,
and a two-mode squeezed vacuum state produced in the process of non-degenerate
parametric down-conversion.
\end{abstract}

\pacs{42.50.Dv, 03.65.Ud, 42.50.Ar}

\maketitle

\section{Introduction}
\label{Sec:Introduction}

Over the years, the field of quantum optics has gained a
justified reputation of a practical testing ground for the
foundations of quantum physics. The broad class of nonclassical
states of optical radiation that can be feasibly generated in a
laboratory has allowed one to demonstrate a variety of quantum
phenomena. Furthermore, progress in the detection techniques has
opened up possibilities to realize a wide range of quantum
measurements on the electromagnetic fields. These advances have
led in particular to two interesting developments. On one hand,
it has become feasible to characterize completely the quantum
state of optical radiation using the methods of so-called
quantum tomography \cite{TomoReview}. On the other hand, the
development of sources of correlated photons based on parametric
down-conversion has resulted in enormous progress in tests of
Bell's inequalities \cite{BellandDownConversion} and more
generally provided practical means to realize experimentally a
number of theoretical proposals in the field of quantum
information processing \cite{DikPDC,KwiatPDC,FuruSoreSCI98}.

Among different representations that can be used to
characterize the quantum state of optical radiation, phase space
quasidistribution functions possess several interesting features
\cite{QOinPhaseSpace}. They are a convenient tool in the
analysis of quantum interference, as well as provide insight
into the classical limit of quantum theory, including the
phenomenon of decoherence \cite{ZureNAT01}. Quasidistribution
functions have played an important role in the field of quantum
state measurement since the first experiments on the
reconstruction of the Wigner function by means of optical
homodyne tomography \cite{SmitBeckPRL93} and also the earlier
work on the determination of the $Q$ function using double
homodyne detection \cite{WalkJMO87}. It was subsequently
realized that standard experimental techniques such as
homodyning and photon counting can be combined into an
alternative, more direct method for measuring quasidistribution
functions \cite{WallVogePRA96,BanaWodkPRL96}. This method
directly provides the value of a quasidistribution function at a
specific point of the phase space. The coordinates of that point
are defined by the amplitude and the phase of an auxiliary
coherent field used for homodyning. The basic idea of the direct
method is to perform photocounting on the signal field
superposed at a highly transmissive beam splitter with a
relatively strong coherent field. With a good approximation,
such a procedure adds a coherent amplitude to the signal field,
while retaining all the quantum fluctuations it initially
contained. The unbalanced homodyning scheme for measuring the
Wigner function has been realized in a proof-of-principle
experiment in Ref.~\cite{BanaRadzPRA99}.

The development of unbalanced homodyning has led to a novel
proposal for testing Bell's inequalities in optical systems
\cite{BanaWodkPRL99}. The proposal was based on an observation
that the two-mode generalization of the direct scheme for
measuring quasidistribution functions could be straightforwardly
reinterpreted as a standard arrangement for measuring
correlation functions between two spatially separated
apparatuses satisfying the locality conditions.  The role of
dichotomic variables in this proposal is played either by the
parity of the registered number of photons, or by the presence
of any number of photons. It was demonstrated theoretically for
a selection of input states that such correlation functions are
capable of violating Bell's inequalities
\cite{BanaWodkPRL99,BanaWodkPRA98}. 
In addition, the character of measured
observables does not require supplementary assumptions that were
necessary in the previous proposals for testing Bell's inequalities
using balanced homodyning \cite{GranPotaYurkPRA88}.
This has opened up yet another
route to optical tests of Bell's inequalities, alternative to
those based on polarization \cite{BellandDownConversion} or
frequency-time entanglement \cite{FrequencyBell}. Recently, the
possibility of testing Bell's inequalities based on homodyning
and photon counting has been explored experimentally by Kuzmich
{\em et al.} \cite{KuzmWalmPRL00}.

The purpose of the present paper is two-fold. In the first part, we provide a
complete multimode theory of the direct method for measuring quasidistribution
functions. Previous theoretical discussions of unbalanced homodyning were based
on an assumption that the beams interfered at a beam splitter contained single
excited modes with perfectly matching spatio-temporal characteristics. Here we
will analyze the case when both the beams are described by general
electromagnetic field operators. We will demonstrate that if the signal field
comprizes several excited modes, the full multimode quasidistribution function
of the signal field can be measured directly by an appropriate spatio-temporal
shaping of the auxiliary coherent field used for homodyning. Interestingly,
this measurement scheme does not depend on whether the excited modes can be
separated before the detection or not. The only observable that needs to be
determined is the statistics of the total number of photons in all the modes of
interest. The general multimode treatment will also allow us to address
rigorously the problem of imperfect mode matching between the signal and probe
fields. We will show that the effect of mode mismatch is essentially different
from that occurring in balanced homodyne detection \cite{BHDMismatch}, where it
simply contributes to the overall detection efficiency. For completeness, we
will also discuss here other practical aspects that are likely to affect a
realistic experiment, such as dark counts.

In the second part of this paper, we apply the multimode theory to perform a
feasibility study of tests of Bell's inequalities based on unbalanced
homodyning. We focus our attention on two quantum states of optical radiation:
a single photon divided by a 50:50 beam splitter and a two-mode squeezed vacuum
state. Both these states can be practically generated using spontaneous
parametric down-conversion in media with $\chi^{(2)}$ nonlinearity. We
determine here the range of experimental parameters such as detector efficiency
and mode matching, which enable violation of Bell's inequalities. Generation of
a one photon Fock state in a controlled spatio-temporal mode has been recently
demonstrated by Lvovsky {\em et al.}\ \cite{LvovHansPRL01}, who were able to
achieve a sufficient overlap with the local oscillator field to demonstrate the
negativity of the corresponding Wigner function. This significant experimental
result encourages us to examine more closely the possibility of testing Bell's
inequality by homodyning a single photon divided on a beam splitter. Generation
of the second state studied here, i.e.\ the two mode squeezed vacuum state,
also attracts presently a great deal of interest as one of the constructional
primitives in quantum information processing based on continuous variables
\cite{FuruSoreSCI98,ContinuousQIP}. We note that in contrast to a typical
scenario considered in this context, involving measurements of continuous
quadrature observables, we will be dealing with a discrete, binary detection
scheme. This example suggests that continuous-variable entanglement could
perhaps be useful also for implementations of other quantum information
protocols relying on binary logic \cite{ChenPanPRL02}.

We would like to stress here that the operational meaning of two-mode
quasidistribution functions as nonlocal correlation functions depends
critically on the specific experimental scheme used for the measurement. In
order to perform a valid test of quantum nonlocality, the measured quantities
must satisfy the local reality conditions that are assumed in the derivation of
the corresponding Bell's inequality. In particular, if standard Bell's
inequalities derived for binary measurements on spin-1/2 particles are to be
applied to a quantum optical experiment, the output signals from the
photodetectors have to be converted into dichotomic variables on a shot-by-shot
basis. This condition is fulfilled by the direct scheme based on unbalanced
homodyning, where in each run one either registers the photon number parity, or
checks for the presence of any number of photons. In contrast, this is not the
case for the optical homodyne tomography scheme, where the Wigner function is
reconstructed by an application of tomographic back-projection algorithms to
quadrature statistics obtained from balanced homodyne detection. Obviously,
such a procedure cannot be performed on the shot-by-shot basis. This difference
can be best seen on the example of the two-mode squeezed vacuum state: it is
well known that detection of quadratures has a straightforward local hidden
variable model based on the Wigner representation, whereas it has been
demonstrated that unbalanced homodyning can reveal the nonlocality of this
state. The above argument invalidates the claim of Ref.~\cite{WuPRA00} that in
order to improve the detection efficiency, unbalanced homodyning can be simply
replaced in tests of Bell's inequalities by optical homodyne tomography
\cite{SmitBeckPRL93} or cascaded homodyning \cite{KisKissPRA99}. We note that
balanced homodyne detection can be used to test Bell's inequalities in other
proposals, that are for example based on a legitimate procedure of binary
classification of the result of a quadrature measurement
\cite{GilcDeuaPRL98,DragBanaPRA01}.

This paper is organized as follows. We start in Sec.~\ref{Sec:MultimodeTheory}
from analyzing multimode interference at an unbalanced beam splitter, and
demonstrate, using an appropriate modal decomposition, how the overall
statistics of photocounts can be related to multimode quasidistribution
functions of the input signal field. In Sec.~\ref{Sec:Practicalities} we
discuss effects of imperfections in a practical experimental setup, including
non-unit detection efficiency, dark counts, and imperfect mode overlap. We find
that these three factors have distinctively different effects on the measured
observables. Next, we review in Sec.~\ref{Sec:Testing} how the two-mode
case can be used to test Bell's inequalities, and derive realistic
expressions for measured correlation functions that include effects of dominant
experimental imperfections. These correlation functions are analyzed in detail
in the following sections using two specific examples of two-mode input states:
a one-photon Fock state divided on a 50:50 beam splitter in
Sec.~\ref{Sec:SinglePhotonState}, and a two-mode squeezed vacuum state in
Sec.~\ref{Sec:TwoModeSqueezedVacuum}. In particular, we find the range of
experimental parameters where a violation of Bell's inequalities can be
observed. Finally, Sec.~\ref{Sec:Conclusions} concludes the paper.

\section{Multimode theory}
\label{Sec:MultimodeTheory}

In this section we develop the full multimode theory of the
direct scheme for measuring the quantum optical
quasidistribution functions proposed in
Refs.~\cite{WallVogePRA96,BanaWodkPRL96}. A general arrangement
depicted in Fig.~\ref{Fig:General} consists of two fields, named
as the signal and the probe, interfered at a beam splitter
characterized by a power transmission $T$. The output port of
the beam splitter is monitored by a photon counting detector
integrating the incident light over its active surface. In the
single-mode description of light incident on a beam splitter it
is sufficient to use a pair of annihilation operators referring
to the signal and the probe fields, respectively. The
spatio-temporal characteristics of these two modes are not
important in this approach as long as it is implicitly assumed
that the modes are matched perfectly at the beam splitter. We
shall free our further analysis from these simplifying
assumptions.

\subsection{Interference at a beam splitter}

Let us denote by $\hat{\bf E}^{(+)}_{\text{\scriptsize out}}({\bf r},t)$ the
positive-frequency part of the electric field operator at the surface of the
detector. This field is a superposition of the signal and the probe fields
combined at the beam splitter BS. Mathematical representation of this
combination is a slightly intricate matter. If we wanted to proceed rigorously
and express $\hat{\bf E}_{\text{\scriptsize out}}^{(+)}({\bf r},t)$ in terms of
the signal and probe field operators before the beam splitter, we would have to
introduce appropriate propagators for the fields in the presence of the beam
splitter \cite{QO-VWW}. Instead, we shall choose different notation for the
signal and the probe fields, which will make the discussion more intuitive. Let
us denote by $\hat{\bf E}_{\text{S}}^{(+)}({\bf r}, t)$ the electric field
operator of the signal beam that would fall onto the detector surface {\em in
the absence of the beam splitter BS}. Analogously, let $\hat{\bf
E}_{\text{P}}^{(+)}({\bf r}, t)$ be the probe field at the detector surface,
assuming that the beam splitter BS {\em was replaced by a perfectly reflecting
mirror}. With these definitions, the field $\hat{\bf E}^{(+)}_{
\text{\scriptsize out}}({\bf r}, t)$ resulting from the interference of the
signal and the probe beams is given simply by
\begin{equation}
\hat{\bf E}^{(+)}_{\text{\scriptsize out}}({\bf r}, t) = \sqrt{T}
\hat{\bf E}^{(+)}_{\text{S}}({\bf r}, t) - \sqrt{1-T} \hat{\bf
E}^{(+)}_{\text{P}}({\bf r}, t),
\end{equation}
where $T$ denotes the beam splitter power transmission coefficient. We have
assumed here that the characteristics of the beam splitter are constant over
the spectral and polarization range of the considered fields. Further, we shall
assume that the detected fields are quasi-monochromatic with the central
frequency $\omega_0$. This will allow us to relate easily the number of photons
to the energy of the field absorbed by the detector. Assuming that the
direction of propagation of the field $\hat{\bf E}^{(+)}_{ \text{\scriptsize
out}}({\bf r}, t)$ is perpendicular to the detector, the operator of the photon
flux through the detector surface reads:
\begin{equation}
\hat{\cal J}_{\text{\scriptsize out}} =
\frac{2\epsilon_0 c}{\hbar\omega_0}
\int_{\Delta t} \text{d}t
\int_{D} \text{d}^2 {\bf r} \,
\hat{\bf E}^{(-)}_{\text{\scriptsize out}}({\bf r}, t)
\hat{\bf E}^{(+)}_{\text{\scriptsize out}}({\bf r}, t),
\end{equation}
where $\hat{\bf E}^{(-)}_{\text{out}}({\bf r},
t) =[\hat{\bf E}^{(+)}_{\text{out}}({\bf r},
t)]^\dagger$, and the temporal and the spatial integrals
are performed respectively over the detector opening time
$\Delta t$ and its active surface $D$. The probability $p_n$
of registering $n$ counts on a photodetector is given by
\begin{equation}
\label{Eq:MultimodeCountStatistics}
p_n = \left\langle : e^{-\eta \hat{\cal
J}_{\text{out}}} \frac{(\eta \hat{\cal
J}_{\text{\scriptsize out}})^n}{n!} : \right
\rangle_{\text{S,P}},
\end{equation}
where $\eta$ is the detector quantum efficiency and $:\ldots:$ denotes
normal ordering.
Following the idea of Ref.~\cite{BanaWodkPRL96}, we shall use
the measured photocount statistics to evaluate the expression
of the form
\begin{equation}
\label{Eq:Pis} \Pi(s) = \sum_{n=0}^{\infty} \left( \frac{s+1}{s-1} \right)^{n}
p_n,
\end{equation}
where $s$ is a real parameter, assumed to be nonpositive in order
to assure convergence of the series on the right-hand side of the above
formula.
We will see that in the limit of negligible losses,
the parameter $s$ describes the
ordering of the measured quasidistribution function.
The quantity in Eq.~(\ref{Eq:Pis}) can be written in terms of the photon
flux operator as:
\begin{equation}
\label{parity}
\Pi(s) =
\left\langle:\!\exp\left(
-\frac{2 \eta \hat{\cal J}_{\text{\scriptsize out}}}{1-s}
\right)\!:\right\rangle_{\text{S,P}}.
\end{equation}
If a coherent field is used as the probe, we may immediately
evaluate the quantum expectation value over the probe field and
obtain:
\begin{widetext}
\begin{eqnarray}
\label{Eq:Sum=ExpInt} \Pi(s) & = & \left\langle  : \exp \left( - \frac{4\eta
\epsilon_0 c}{\hbar\omega_0 (1-s)} \int_{\Delta t} \text{d}t \int_{D}
\text{d}^2 {\bf r} \, \left[\sqrt{T} \hat{\bf E}_{\text{S}}^{(-)}({\bf r}, t) -
\sqrt{1-T} {\bf E}_{\text{P}}^\ast({\bf r}, t)\right]
\right. \right.
\nonumber \\
& & \times \left. \left.
\vphantom{\frac{4\eta
\epsilon_0 c}{\hbar\omega_0 (1-s)} \int_{\Delta t} \text{d}t \int_{D}
\text{d}^2 {\bf r}}
\left[\sqrt{T} \hat{\bf
E}^{(+)}_{\text{S}}({\bf r}, t) - \sqrt{1-T} {\bf E}_{\text{P}}({\bf r},
t)\right] \right) : \right\rangle,
\end{eqnarray}
\end{widetext}
where ${\bf E}_{\text{P}}({\bf r}, t) = \langle
\hat{\bf E}^{(+)}_{\text{P}}({\bf r}, t)
\rangle_{\text{P}}$ is the amplitude of the electric field of
the probe beam. Thus, we can simply replace the coherent probe field
operators with their expectation values in all normally ordered
expressions involving the photocount statistics.

\subsection{Modal decomposition}

We will now consider the signal field $\hat{\bf
E}^{(+)}_{\text{S}}({\bf r}, t)$ in which a finite
number of $M$ modes is possibly excited. We shall denote the
corresponding annihilation operators by $\hat{a}_i$, and the
mode functions by ${\bf u}_i ({\bf r}, t)$, where
$i=1,2,\ldots M$. Our goal will be to relate the photon
statistics $p_n$ to the multimode quasidistribution
characterizing these modes. Thus we decompose the signal
field $\hat{\bf E}^{(+)}_{\text{S}}({\bf r}, t)$ in
the form
\begin{equation}
\label{Eq:ESDecomposition}
\hat{\bf E}^{(+)}_{\text{
S}}({\bf r}, t) = \sum_{i=1}^{M} \hat{a}_i {\bf u}_i ({\bf
r}, t) + \hat{\bf V}({\bf r}, t),
\end{equation}
where the operator $\hat{\bf V}({\bf r}, t)$ is a sum of
all the other modes remaining in the vacuum state. This
part of the field does not contribute to the detector
counts in the normally ordered expression given in
Eq.~(\ref{Eq:Sum=ExpInt}).

Further, we shall assume that virtually all the excited
part of the signal field is absorbed by the detector within
the gate opening time.  This allows us to write
orthonormality relations for the mode functions ${\bf
u}_i({\bf r}, t)$ in the form
\begin{equation}
\label{Eq:uOrthonormality}
\frac{2\epsilon_0 c}{\hbar\omega_0}
\int_{\Delta t} \text{d}t
\int_{D} \text{d}^2 {\bf r} \,
{\bf u}_i^\ast ({\bf r}, t){\bf u}_j ({\bf r}, t) =
\delta_{ij},
\end{equation}
where the integrals are restricted to the domain defined by
the detection process.
With these assumptions, we may simplify the exponent of
Eq.~(\ref{Eq:Sum=ExpInt}). It is convenient to introduce
dimensionless amplitudes $\alpha_i$, which are projections
of the probe field onto the signal mode functions:
\begin{equation}
\label{Eq:alphaidef}
\alpha_i = \frac{2\epsilon_0 c}{\hbar\omega_0}
\int_{\Delta t} \text{d}t
\int_{D} \text{d}^2 {\bf r} \,
{\bf u}_i^\ast ({\bf r}, t)
{\bf E}_{\text{P}}({\bf r}, t).
\end{equation}
These amplitudes describe the parts of the coherent probe field
${\bf E}_{\text{P}}({\bf r}, t)$ that match the corresponding modes
in the signal field. Furthermore, let us also denote by
${\cal J}_{\text{P}}$ the average total number of photons in the
probe field:
\begin{equation}
{\cal J}_{\text{P}} =
\frac{2 \epsilon_0 c}{\hbar\omega_0}
\int_{\Delta t} \text{d}t \int_{D} \text{d}^2 {\bf r} \,
|{\bf E}_{\text{P}}({\bf r}, t)|^2.
\end{equation}
Using this notation, we may write the quantity $\Pi(s)$ determined
from the photocount statistics as:
\begin{widetext}
\begin{eqnarray}
\label{summ}
\Pi(s)
& = & 
\left\langle
: \exp\left[- \frac{2\eta T}{1-s}
\sum_{i=1}^{M} \left(\hat{a}_i^\dagger -
\sqrt{\frac{1-T}{T}}
\alpha_i^\ast \right)\left(\hat{a}_i -
\sqrt{\frac{1-T}{T}}
\alpha_i \right) \right] :  \right\rangle
\nonumber \\
& & \times
\exp \left[
 - \frac{2\eta(1-T)}{1-s}
\left({\cal J}_{\text{P}} - \sum_{i=1}^{M}
|\alpha_i|^2 \right) \right].
\end{eqnarray}
The expectation value appearing in the above expression can be identified, up
to a normalization factor, as an $M$-mode quasidistribution function. This
equivalence becomes clear if we recall the normally ordered definition of
the generalized $s$-ordered quasidistribution functions \cite{CahiGlauPR69}:
\begin{equation}
W(\gamma_1, \ldots , \gamma_M ; s ) = \left(\frac{2}{\pi(1-s)}\right)^M
\left\langle : \! \exp \left( - \frac{2}{1-s} \sum_{i=1}^{M} (\hat{a}_i^\dagger
- \gamma_i^\ast)(\hat{a}_i - \gamma_i ) \right) \! : \right\rangle.
\end{equation}
After the identification of the parameters we obtain that:
\begin{eqnarray}
\Pi(s)
& = &
\left(
\frac{\pi (1-s)}{2\eta T} \right) ^{M}
W_{\text{S}} \left(
\sqrt{\frac{1-T}{T}}\alpha_1, \ldots,
\sqrt{\frac{1-T}{T}}\alpha_M
;
- \frac{1-s- \eta T}{\eta T}
\right)
\nonumber \\
& & \times
 \exp \left[
 - \frac{2\eta(1-T)}{1-s}
\left({\cal J}_{\text{P}} - \sum_{i=1}^{M}
|\alpha_i|^2 \right) \right].
\end{eqnarray}
\end{widetext}
The exponential factor multiplying the quasidistribution function
involves the difference between the average total number of photons
$\hat{\cal J}_{P}$ in the probe field and the sum $\sum_{i=1}^{M}
|\alpha_i|^2$ describing the number of the probe field photons that
match the corresponding modes of the signal field. This factor results
from the part of the probe field that is orthogonal (in the sense of
spatio-temporal overlap defined by Eq.~(\ref{Eq:uOrthonormality})) to the
mode functions describing the excited component of the signal beam, and
it becomes identically equal to one if the probe field exactly matches
the $M$ signal modes of interest. The matching condition can be written as:
\begin{equation}
\label{Eq:EPPerfectMatching}
{\bf E}_{\text{P}}({\bf r}, t) = \sum_{i=1}^{M}
\alpha_i {\bf u}_i ({\bf r}, t).
\end{equation}
In such a case of perfect mode matching on the beam splitter the quantity
$\Pi(s)$ calculated from the photocount statistics is equal, up to a constant
multiplicative factor, to the value of a multimode quasidistribution function with the
ordering $-(1-s-\eta T)/\eta T$, taken at the point $(\sqrt{(1-T)/T} \alpha_1 ,
\ldots , \sqrt{(1-T)/T} \alpha_M)$. This result implies a simple recipe for
measuring multimode quasidistribution functions point-by-point. What one needs
to do in order to determine the value of a quasidistribution function at a
specific point, is to prepare the coherent probe field in an appropriate
superposition defined by Eq.~(\ref{Eq:EPPerfectMatching}), and to measure the
photocount statistics of the probe and signal fields combined at an unbalanced
beam splitter. The sum over the photocount statistics evaluated according to
Eq.~(\ref{Eq:Pis}) yields directly the value of the quasidistribution function.
The ordering of the measured quasidistribution depends on the overall losses of
the signal field characterized by the product $\eta T$, as well as the
parameter $s$ used to evaluate the sum over the count statistics. The allowed
range of the parameter $s$ has been discussed previously \cite{BanaWodkJMO97}
in the context of the statistical uncertainty. It has been shown that in order
to keep the statistical error bounded, the parameter $s$ needs to be taken
nonpositive. As in the single-mode case, the highest attainable ordering of the
measured quasidistribution function is equal to $-(1-\eta T)/\eta T$, which in
the limit of ideal lossless detection yields zero, thus corresponding to the
detection of the Wigner function.  An interesting feature of this scheme is
that there is no need to resolve contributions to the photocount statistics
from each of the modes separately; the only observable that needs to be
reconstructed is the sum over the statistics of the total number of
photocounts.

One should note that in a realistic situation it is not always possible
to measure the complete photocount statistics. Most of the single-photon
detectors used presently, such as avalanche photodiodes operated in the
Geiger mode, yield only a binary response telling whether any number of
photons has been registered or not. We can easily specialize our previous
considerations to include this case by taking the limit $s \rightarrow
-1$, in which the sum $\Pi(s)$ reduces to
\begin{equation}
\lim_{s \rightarrow -1} \Pi(s) = p_0,
\end{equation}
i.e.\ the probability of registering zero counts on the detectors.
Thus, if only the information about the presence or absence of photons
is available, we can reconstruct a quasidistribution function with the
ordering $-(2-\eta T)/\eta T$, which in the limit of lossless detection
reduces to $-1$ corresponding to the measurement of a $Q$ function.

More generally, one can consider application of a detection scheme
that can resolve the number of photons up to a certain value, such as
the visible light photon counter \cite{KimTakeAPL99},
detector cascading \cite{KokBrauPRA01}, or the loop detector \cite{BanaWalmXXX02}.
In this case, one is limited to sampling these regions of the phase space
where the photon statistics effectively vanishes above the cut-off value
of the used detector. Several examples of photon statistics have been analyzed
in detail in Ref.~\cite{BanaWodkJMO97}. It has been found in certain cases
that the occurence
of structures in the phase space can be related to destructive interference
at the unbalanced beam splitter, resulting in low photon numbers seen by the detectors.
These structures could thus be measured using detectors with restricted multiphoton resolution.

\subsection{Multiple detectors}
\label{Sec:ManyDetectors}

The direct scheme for measuring multimode quasidistribution
functions can be easily extended to the case when some of the
modes are separated in space. In this case, each mode (or a group
of spatially overlapping modes) needs to be displaced in the phase
space by combining it at a beam splitter with a coherent probe
field, and then measured using a photon counting detector. For
simplicity, let us assume that we are dealing with two separate
beams labelled with $A$ and $B$.  Such a setting, with the
detection procedures satisfying locality conditions, will serve as
a scheme for testing Bell's inequalities discussed in the
subsequent sections of this paper. Generalization of our results
to more that two beams will follow in a straightforward manner.
The two detectors in our setup yield photocount statistics
$p_{A,n}$ and $p_{B,n}$. We will show that Eq.~(\ref{Eq:Pis})
applied to the statistics of the total number of counts $p_n$,
defined as:
\begin{equation}
p_n = \sum_{m=0}^{n} p_{A,m} p_{B,n-m}
\end{equation}
yields the joint quasidistribution function of the signal field.

In order to fix the notation, let us assume that $M_A$ and $M_B$ modes are
excited in the beams $A$ and $B$ respectively, and denote the corresponding
annihilation operators by $\hat{a}_i$ and $\hat{b}_i$. The displacement of the
signal field can be performed by preparing the probe field in the form of two
beams interfering with the beams $A$ and $B$. A calculation completely
analogous to the one presented in the previous section shows that $\Pi(s)$
evaluated from the overall count statistics has the following form:
\begin{widetext}
\begin{eqnarray}
\Pi (s) &=&
\left\langle
:\!\exp\left(- \frac{2\eta T}{1-s}
\sum_{i=1}^{M_A}
\left(\hat{a}_i^\dagger -
\sqrt{(1-T)/T}
\alpha_i^\ast \right)
\left(\hat{a}_i -
\sqrt{(1-T)/T}
\alpha_i\right)
\right) \right.
\nonumber \\
& &
\left.
\otimes
\exp\left(- \frac{2\eta T}{1-s}
\sum_{i=1}^{M_B}
\left(\hat{b}_i^\dagger -
\sqrt{(1-T)/T}
\beta_i^\ast \right)
\left(\hat{b}_i -
\sqrt{(1-T)/T}
\beta_i\right) \right)\!:
\right\rangle
\nonumber \\
\label{Eq:summ2}
& & \times \exp \left[- \frac{2\eta(1-T)}{1-s}
\left({\cal J}_{\text{P}, A}
- \sum_{i=1}^{M_A}
|\alpha_i|^2
+ {\cal J}_{\text{P}, B}
- \sum_{i=1}^{M_B}
|\beta_i|^2 \right)
\right]
\end{eqnarray}
\end{widetext}
Here $\alpha_i$ and $\beta_i$ are the probe field amplitudes
matching the modes of the signal field. They are defined
analogously to Eq.~(\ref{Eq:alphaidef}), with the integration
over the detection time and the active area of the appropriate
detector $A$ or $B$. Similarly, ${\cal J}_{\text{P}, A}$
and ${\cal J}_{\text{P}, B}$ are the intensities of the two
probe beams expressed in the photon number units.
For simplicity we have assumed here that the two detectors
have the same quantum efficiency $\eta$, and also that both the beam
splitters are characterized by the same transmission coefficient $T$.
As before, the expectation value in Eq.~(\ref{Eq:summ2}) can be
identified as proportional to a multimode quasidistribution
function, and the second exponential factor results from imperfect
matching of the probe field to the signal modes.

\section{Practicalities}
\label{Sec:Practicalities}

In this section, we will discuss the effect of experimental
imperfections that are likely to occur in a practical realization
of the proposed scheme. First, we shall analyze the role of dark
counts in the described measurement scheme. Secondly, we will
study on a simple example the effect of imperfect matching between
the signal and probe fields at the beam splitter.

\subsection{Dark counts}
\label{SubSec:DarkCounts}

The first effect that needs to be taken into account is the dark noise of
the detectors. Apart from the photocounts originating from the measured
field, the detector can generate additional counts. These can result from
thermal excitations in the active material of the photodetector, or from
absorption of stray light. We shall make a simplifying assumption in
our analysis that the dark counts are not correlated statistically
with the photocounts generated by the field of interest.
Thus we neglect for example the phenomenon of afterpulsing in avalanche
photodiodes, which consists in generating signal by carriers
trapped after detection of preceding photons.
Under this
assumption the statistics of detector counts $p_n$ can be represented
as a discrete convolution of the statistics of counts originating from
the measured field $p_{{\text{F}},n}$ and the statistics of dark counts
$p_{{\text{D}},n}$:
\begin{equation}
\label{Eq:pnDarkCounts}
p_n = \sum_{m=0}^{n} p_{{\text{F}},m} p_{{\text{D}},n-m}.
\end{equation}
It is then straightforward to check that the sum over the count
statistics factorizes into the product:
\begin{eqnarray}
\sum_{n=0}^{\infty}
\left( \frac{s+1}{s-1} \right)^{n}
p_n
& = &
\sum_{m=0}^{\infty}
\left( \frac{s+1}{s-1} \right)^{m}
p_{{\text{F}},m}
\nonumber \\
& &
\times
\sum_{n=0}^{\infty}
\left( \frac{s+1}{s-1} \right)^{n}
p_{{\text{D}},n}
\end{eqnarray}
This expression means that as long as the dark counts are uncorrelated with the
counts generated by the photons from the measured fields, $\Pi(s)$ simply
becomes rescaled by a constant factor defined only by the dark count
statistics. Therefore in order to include the effect of dark counts in our
analysis we only need to multiply the previous formulas by this constant
factor. Furthermore, if we choose $s=-1$, this multiplicative factor simply
becomes $p_{{\text{D}},0}$, i.e.\ the probability of a zero dark count event.

\subsection{Mode mismatch}
\label{SubSec:ModeMismatch}

In a realistic situation, it is virtually impossible to achieve
ideal matching between the signal and probe fields at a beam
splitter. Consequently, experimental results will usually be
affected by imperfect mode matching. In order to analyze this
effect in quantitative terms and estimate its importance, we will
now consider the practical measurement of a single
quasidistribution function. For this purpose, we will assume that
$M=1$, i.e.\ that only one mode is excited in the decomposition of
the signal field defined in Eq.~(\ref{Eq:ESDecomposition}), and
for simplicity we shall drop the index labelling its annihilation
operator $\hat{a}$ and the corresponding mode function ${\bf
u}({\bf r}, t)$. In the further analysis, it will be convenient to
introduce the mode matching parameter $\xi$ defined by:
\begin{equation}
\xi = \frac{|\alpha|^2}{{\cal J}_{\text{P}}} = \frac{2 \epsilon_0
c}{\hbar \omega_0} \frac{\displaystyle \left| \int_{\Delta t}
\text{d}t \int_{D}\text{d}^2 {\bf r}
\, {\bf u}^{\ast}({\bf r}, t) {\bf E}_{\text{P}} ({\bf r}, t) \right|^2}%
{\displaystyle \int_{\Delta t} \text{d}t \int_{D} \text{d}^2 {\bf
r} \, | {\bf E}_{\text{P}} ({\bf r}, t) |^2}.
\end{equation}
It follows directly from the Schwarz inequality that
the parameter $\xi$ defined in the above way lies between
$0$ and $1$. The value $\xi =1$ corresponds to the perfect overlap
between the signal and the probe modes, whereas the value $\xi = 0$
means that the signal and probe fields are completely orthogonal.

Using the mode matching parameter, we can write $\Pi(s)$ as:
\begin{eqnarray}
\Pi(s)  & = & \frac{\pi (1-s)}{2 \eta T} W \left(
\sqrt{\frac{1-T}{T}} \alpha ; - \frac{1-s - \eta T}{\eta T}
\right)
\nonumber \\
\label{Eq:PisModeMismatch} & & \times \exp \left( - \frac{2 \eta (1-T)}{1-s}
\frac{1-\xi}{\xi} |\alpha|^2 \right).
\end{eqnarray}
It is seen that in the case of imperfect mode matching the
measured quasidistribution function is multiplied by a Gaussian
envelope centered at the origin of the phase space. The width of
the Gaussian depends on the factor $(1-\xi)/\xi$, which grows
with the increasing mismatch between the signal and probe
fields. Thus, the effect of the mode mismatch is more severe in
the outer regions of the phase space, where the amplitude of the
signal quasidistribution function is suppressed by the Gaussian
envelope resulting from the mode mismatch.

It is interesting to note that the effect of the mode mismatch in
our scheme is quite different from that occurring in balanced
homodyne detection \cite{BHDMismatch}. In the latter case, the
mode matching parameter simply multiplies the overall efficiency
of the homodyne detector, and is equivalent to additional losses
of the signal field. In contrast, Eq.~(\ref{Eq:PisModeMismatch})
includes the quantum efficiency and the mode matching parameter in
two nonequivalent ways, and the strength of the effect of the mode
mismatch depends on the selected point of the phase space. This
difference between balanced and unbalanced homodyning can be
easily understood by comparing the relative intensities of the
signal and local oscillator fields at the detectors. In the balanced case, only the
part of the signal field matching the shape of the local
oscillator affects noticeably the photocount statistics. In the
case of unbalanced homodyning, both the signal and local
oscillator fields have comparable intensities at the exit of the
superposing beam splitter and contribute with similar strengths to
the count statistics.

In a practical situation, it is convenient to relate the mode
matching parameter $\xi$ to the interference visibility
on the beam splitter used to combine the signal and probe
fields. Suppose that we can replace the signal field of interest
by a coherent state $|\alpha_{\text{S}}\rangle$ prepared in
exactly the same spatio-temporal mode. A simple calculation shows
that the average light intensity ${\cal J}_{\text{out}}$ received
by the detector, expressed in the photon number units, is given
by:
\begin{equation}
{\cal J}_{\text{out}} = (1-\xi) T |\alpha_{\text{S}}|^2 + | \sqrt{\xi T}
\alpha_{\text{S}} - \sqrt{(1-T)\xi} \alpha|^2.
\end{equation}
The minimum intensity of the output light that can be obtained by
varying the phase and the amplitude of the probe field while
keeping $\alpha_{\text{S}}$ fixed is given by:
\begin{equation}
{\cal J}_{\text{min}} = (1-\xi) T | \alpha_{\text{S}} |^2
\end{equation}
If we now change the phase of the probe field by $\pi$, this will
maximize the output intensity for the fixed absolute value of the probe field
amplitude,
thus yielding:
\begin{equation}
{\cal J}_{\text{max}} =  (1+ 3 \xi) T |\alpha_{\text{S}}|^2.
\end{equation}
Consequently, the interference visibility $\upsilon$ defined in the
standard way can be expressed in terms of the parameters used thorough
our discussion as:
\begin{equation}
\upsilon
= \frac{{\cal J}_{\text{max}} - {\cal J}_{\text{min}}}%
{{\cal J}_{\text{max}} + {\cal J}_{\text{min}}}
= \frac{2\xi}{1 + \xi}.
\end{equation}
Inverting this relation we obtain that $\xi = \upsilon / (2 -
\upsilon)$. Thus, the measurement of the interference
visibility based on the method sketched above can yield an
estimate for the quality of the mode matching in the
experimental setup. This approach has been used in
Ref.~\cite{BanaRadzPRA99} to analyze the shape of the
experimentally measured Wigner function of a coherent state.

\section{Testing Bell's inequalities}
\label{Sec:Testing}

We will now specialize the general theory developed in the
previous sections to the case when two light modes in an entangled
state are sent towards a pair of spatially separated unbalanced
homodyne detectors. It has been suggested in previous theoretical
works \cite{BanaWodkPRL99,BanaWodkPRA98} that such a setup can be
used to test Bell's inequalities. In the proposed approach, the standard
spin measurements are replaced by dichotomic observables derived
from the photocount statistics, and the auxiliary coherent fields
provide adjustable parameters that can be selected by the
observers under locality conditions. Our goal here will be to
perform a complete realistic analysis of the proposed scheme,
including all the practical factors that are likely to affect the
outcome of the proposed experiment.

Let us first summarize briefly the unbalanced homodyne setup
proposed to test Bell's inequalities, depicted in
Fig.~\ref{Fig:TwoMode}. Two entangled and spatially separated
light beams fall onto unbalanced homodyne detectors composed of
photon counters preceded by high-transmission beam splitters with
auxiliary coherent fields entering through the sideway input
ports. The operation of such a setup can be easily reformulated in
terms of a standard scheme for testing Bell's inequalities with
measuring apparatuses that provide dichotomic outcomes. The binary
variable we will be interested in is whether {\em any} number of
photons has been registered by a photon counting detector,
without resolving the actual number of photons. Such a measurement
corresponds formally to a standard test whether a particle has
passed through a polarization analyzer. It is important to note
that this dichotomic measurement can be feasibly realized with the
help of avalanche photodiodes operated in the Geiger mode that are
not capable of providing the number of photons triggering the
avalanche signal. At the same time, we can utilize other favorable
characteristics of avalanche photodiodes, such as relatively high
quantum efficiency and low dark count rates.
The pair of non-commuting observables at each
receiving station will be established by selecting two different
complex amplitudes of the coherent field added to the signal beam
at the beam splitter. The choice of the coherent state amplitudes
thus corresponds to two possible orientations of polarization
analyzers used in standard experiments on Bell's inequalities. In
each experimental run, the observers select randomly the
amplitudes of the coherent fields and record the response of their
photon counting detectors. In practice, all the coherent light beams
in the setup need to be derived from the same source in order to ensure fixed
phase relationships. This of course fully complies with the assumptions
underlying Bell's inequalities as long as the amplitude and the phase
of auxiliary coherent beams are adjusted by the two parties under
locality conditions. We will now discuss in detail
correlation functions constructed from these data and analyze
their nonlocal character.

In order to formulate our discussion in terms of quasidistribution
functions and utilize results derived in
Secs.~\ref{Sec:MultimodeTheory} and \ref{Sec:Practicalities}, we
shall attribute the value $+1$ to events a photon counting
detector did not click and $0$ when any non-zero number of photons
was registered. In such a case, we can specialize the results from
the previous sections by taking the value of the $s$ parameter
used in the definition of the sum over the photon count statistics
in Eq.~(\ref{Eq:Pis}) equal to $s=-1$. Then the whole sum
over the photon count statistics reduces simply to $p_0$, i.e.\
the probability of observing zero photons, which corresponds
exactly to our assignment of numerical values to experimental
outcomes. Our test of Bell's inequalities is based on measuring
the joint and marginal probabilities of such ``photon silence''
events, which we shall denote respectively by $Q(\tilde{\alpha},
\tilde{\beta})$ and $Q(\tilde{\alpha})$ or $Q(\tilde{\beta})$.
Here $\tilde{\alpha}$ and $\tilde{\beta}$ are complex amplitudes
of the coherent fields rescaled by a certain factor whose value we
will discuss later. As the probability of a no-count event is
obviously bounded between $0$ and $1$, the nonlocal character of
the measured correlation functions can be tested with the help of
the standard Clauser-Horne inequality \cite{ClauHornPRD74}:
\begin{equation}
\label{Eq:CH}
-1 \le {\cal CH} \le 0,
\end{equation}
where the combination ${\cal CH}$ is constructed with the
help of the formula:
\begin{eqnarray} \label{Eq:CHCombination}
{\cal CH} &=&
 Q(\tilde{\alpha}_1, \tilde{\beta}_1 )
+Q(\tilde{\alpha}_1, \tilde{\beta}_2 )
+Q(\tilde{\alpha}_2, \tilde{\beta}_1 )
-Q(\tilde{\alpha}_2, \tilde{\beta}_2 )
\nonumber \\
& &
-Q(\tilde{\alpha}_1)
-Q(\tilde{\beta}_1).
\end{eqnarray}
Here $\tilde{\alpha}_1, \tilde{\alpha}_2$ and $\tilde{\beta}_1,
\tilde{\beta}_2$ denote the two settings of the coherent fields
used at the respective detectors.

We will now derive expressions for the joint and marginal
probabilities of ``photon silence'' events that include
possible imperfections of the experimental setup. For simplicity,
we shall assume that the excited parts of the fields
travelling to the homodyne detectors can be described
by single-mode annihilation operators $\hat{a}$ and
$\hat{b}$. The expression for the joint zero-count
event is then simply given by Eq.~(\ref{Eq:summ2})
with $s=-1$ and $M_A=M_B=1$.
Following the discussion in
Sec.~\ref{SubSec:ModeMismatch}, it will be
convenient to introduce a mode mismatch parameter defined by:
\begin{equation}
\xi = \frac{|\alpha|^2}{{\cal J}_{\text{P}, A}}
= \frac{|\beta|^2} {{\cal J}_{\text{P}, B}}.
\end{equation}
assumed to be the same in the both arms of the setup. In addition,
we will also include dark counts on the detectors. As we are
interested only in the probability of zero-count events, it
follows from Eq.~(\ref{Eq:pnDarkCounts}) that the only relevant
parameter is $p_{\text{D},0}$, i.e.\ the probability that zero
dark counts have occurred. For brevity, we will denote this
quantity simply as $p_{\text{D}}$. Combining all the above
elements yields the full expression for the probability of a joint
``photon silence'' event:
\begin{widetext}
\begin{eqnarray}
\label{jointsilence} Q(\tilde{\alpha}, \tilde{\beta} ) &=&
p_{\text{D}}^2 \left\langle :\!\exp[- \eta T \left(\hat{a}^\dagger
- \tilde{\alpha}^\ast \right) \left(\hat{a} -
\tilde{\alpha}\right) ]
 \otimes \exp [- \eta T (\hat{b}^\dagger -
\tilde{\beta}^\ast ) (\hat{b} - \tilde{\beta})
]\!: \right\rangle 
\nonumber \\
& & \times \exp \left( - \eta T\frac{1-\xi}{\xi}
(|\tilde{\alpha}|^2 + |\tilde{\beta}|^2 ) \right)
\nonumber \\
&=& \left( \frac{\pi p_{\text{D}}}{\tilde{\eta}} \right)^2 W_{AB}
\left( \tilde{\alpha} , \tilde{\beta} ; - \frac{2-
\tilde{\eta}}{\tilde{\eta}} \right) \exp\left( - \tilde{\eta}
\frac{1-\xi}{\xi} ( |\tilde{\alpha}|^2 + |\tilde{\beta}|^2
)\right),
\end{eqnarray}
\end{widetext}
where $W_{AB} ( \tilde{\alpha} , \tilde{\beta} ; s)$
is the two-mode $s$-ordered quasidistribution function describing
the entangled signal state.
In the above formula, we have introduced several tilded
quantities that will be convenient as independent parameters
in the further discussion.
First, $\tilde{\alpha}=\sqrt{(1-T)/T}\alpha$ and
$\tilde{\beta}=\sqrt{(1-T)/T}\beta$ denote the rescaled
amplitudes of the auxiliary coherent fields. The product
$\tilde{\eta} = \eta T$ describes the combined losses of the
signal field at a beam splitter and a detector, and can be
easily generalized to include also other sources of losses
experienced by the signal beams.
We can find in a similar manner that
the zero-count probability on the
detector $A$ is equal:
\begin{equation}
\label{singlesilence} Q(\tilde{\alpha} ) = \frac{\pi
p_{\text{D}}}{\tilde{\eta}} W_A \left( \tilde{\alpha} ; - \frac{2-
\tilde{\eta}}{\tilde{\eta}}\right) \exp\left( - \tilde{\eta}
\frac{1-\xi}{\xi}
 |\tilde{\alpha}|^2 \right),
\end{equation}
and analogously for the detector $B$. Thus, in general the joint
and marginal ``photon silence'' probabilities are proportional to
corresponding quasidistribution functions that describe the
properties of the entangled state used in the scheme. It is seen
that the three parameters related to different imperfections:
overall efficiency $\tilde{\eta}$, mode mismatch parameter $\xi$,
and zero dark count probability $p_D$ enter these expressions in a
nonequivalent way. As a lower value of any of them means more
imperfections of the experimental setup, one can expect a
universal behavior that the violation of Bell's inequalities
becomes more pronounced with the increasing values of these
parameters. This will be confirmed in all the examples that we
will study later.

In the following two sections, we will discuss two specific
quantum states that can be produced with the help of spontaneous
parametric down-conversion: a one-photon state divided on a 50:50
beam splitter and a two-mode squeezed vacuum state. The states
used in our discussion formally exhibit entanglement when written
in the Fock basis. Being aware of conceptual difficulties related
to this kind of entanglement \cite{GHZ} we shall focus on the
violation of Bell's inequalities as a signature of nonlocality
that can be verified in practice. In order to reveal a violation
of Bell's inequalities, it is necessary to use measurements that
probe the coherence between different Fock terms. In particular,
it is widely recognized that a single particle alone is not
sufficient to demonstrate quantum nonlocality using particle
counting detectors \cite{BiednyLucien}. This is why we need local
supplies of additional photons in the form of auxiliary coherent
fields to detect nonlocal correlation functions. Similarly, for a
two-mode squeezed vacuum state, counting photons alone is not
sufficient to probe the coherence between different Fock states in
the superposition and demonstrate entanglement. We need auxiliary
coherent fields to make the detected observables sensitive to the
off-diagonal elements of the input density matrix in the Fock
basis.

\section{Single photon state}
\label{Sec:SinglePhotonState}

As the first example we will consider a single photon divided at a 50:50 beam
splitter. The single-photon state can be produced in a controlled way with the
help of the spontaneous down-conversion process in a $\chi^{(2)}$ nonlinear
medium assisted by the conditional measurement. Photons generated in this
process are always emitted in pairs towards two well defined directions that
are determined by the phase-matching conditions inside the nonlinear crystal.
One component of such a pair can be used as a trigger yielding a definite
information about emission of the second photon. The latter can be consequently
sent to the 50:50 beam splitter. From the formal point of view the output
two-mode state of the field leaving the beam splitter resembles a single photon
Fock state entangled with a vacuum mode entering through the unused port of the
beam splitter:
\begin{equation}
\label{singlephoton}
|\Psi\rangle = \frac{1}{\sqrt{2}}
\left(|1\rangle_A |0\rangle_B + |0\rangle_A |1\rangle_B
\right).
\end{equation}
The output modes have been denoted here with the indices $A$ and
$B$. We will be interested in correlation functions
(\ref{jointsilence}) and their marginals (\ref{singlesilence}) of
"photon silence" events on the detectors measured as
a function of the amplitudes of
the auxiliary coherent beams. Following Eqs.~(\ref{singlesilence})
and (\ref{jointsilence}) we need to evaluate first
the generalized quasidistribution function of the state
(\ref{singlephoton}) and its one-mode marginal.
After simple algebra we
obtain:
\begin{eqnarray}
\lefteqn{W_{AB} \left( \tilde{\alpha} , \tilde{\beta} ; -
\frac{2- \tilde{\eta}}{\tilde{\eta}} \right)} & &
\nonumber \\
& = &\left(
\frac{\tilde{\eta}}{\pi} \right)^2 \langle\Psi|
:\!e^{ -\tilde{\eta}\left(
(\hat{a}^\dagger-\tilde{\alpha}^*)(\hat{a}-\tilde{\alpha})
+(\hat{b}^\dagger-\tilde{\beta}^*)(\hat{b}-\tilde{\beta})
\right)}\!:
|\Psi\rangle
\nonumber \\
& = & \left( \frac{\tilde{\eta}}{\pi} \right)^2
\left(1-\tilde{\eta}+\frac{\tilde{\eta}^2}{2}
|\tilde{\alpha}+\tilde{\beta}|^2\right)
e^{-\tilde{\eta}\left(|\tilde{\alpha}|^2+|\tilde{\beta}|^2
\right)},
\end{eqnarray}
with the marginal, single-mode quasidistribution:
\begin{eqnarray}
W_{A} \left( \tilde{\alpha} ; - \frac{2-
\tilde{\eta}}{\tilde{\eta}} \right)
& = & \frac{\tilde{\eta}}{\pi} \langle\Psi|
:\!e^{ -\tilde{\eta}(\hat{a}^\dagger-\tilde{\alpha}^*)
(\hat{a}-\tilde{\alpha})}\!:|\Psi\rangle
\nonumber \\
& = & \frac{\tilde{\eta}}{\pi}
\left(1-\frac{\tilde{\eta}}{2}+\frac{\tilde{\eta}^2}{2}
|\tilde{\alpha}|^2\right) e^{-\tilde{\eta}|\tilde{\alpha}|^2}.
\nonumber \\
& &
\end{eqnarray}
Multiplying these expressions by the state-independent factors
given in Eqs.~(\ref{jointsilence}) and (\ref{singlesilence}),
we can easily calculate the joint and marginal probabilities
of the ``photon silence'' events:
\begin{eqnarray}
Q(\tilde{\alpha}, \tilde{\beta} )&=&
p_{\text{D}}^2\left(1-\tilde{\eta}+\frac{\tilde{\eta}^2}{2}
|\tilde{\alpha}+\tilde{\beta}|^2\right)
\nonumber \\
\label{Eq:QABSinglePhoton}
& &\times
\exp\left[
-\frac{\tilde{\eta}}{\xi}\left(|\tilde{\alpha}|^2+
|\tilde{\beta } | ^ 2 \right)\right]
\\
\label{Eq:QASinglePhoton}
Q(\tilde{\alpha})&=& p_{\text{D}}\left(1-\frac{\tilde{\eta}}{2}
+\frac{\tilde{\eta}^2}{2} |\tilde{\alpha}|^2\right)\exp\left(
-\frac{\tilde{\eta}}{\xi}|\tilde{\alpha}|^2\right).
\end{eqnarray}
As a function of $\tilde{\alpha}$ and $\tilde{\beta}$,
these results are the well known hat-shaped distributions.

We will now address the question under what conditions
the correlations between detecting zero-count events
given by the above formulas
lead to the violation of the Clauser-Horne inequality
given in Eq.~(\ref{Eq:CH}). As we have discussed earlier,
there are three non-equivalent
parameters that describe imperfections
of the experimental setup: the overall detection efficiency
$\tilde{\eta}$, the mode-matching parameter $\xi$, and the
zero dark count rate $p_D$. We would like to identify
the range of these parameters that enables one to observe
a violation of Bell's inequalities using unbalanced
homodyning. The parameters that can be freely adjusted in
a practical setup are the values of the four coherent displacements
$\tilde{\alpha}_1, \tilde{\alpha}_2, \tilde{\beta}_1$,
and $\tilde{\beta}_2$ used to construct the Clauser-Horne
combination given in Eq.~(\ref{Eq:CHCombination}). We
shall therefore optimize them for a specified set of
$\tilde{\eta}, \xi$, and $p_D$ in order to find the maximum
possible violation of the Clauser-Horne under given experimental
constraints.

The task of finding the optimal quadruplet of coherent
displacements is a nonlinear multidimensional optimization
problem and we have approached it numerically with the help of the
standard downhill simplex algorithm described in
Ref.~\cite{Amoeba}. It is obvious from the form of
Eqs.~(\ref{Eq:QABSinglePhoton}) and (\ref{Eq:QASinglePhoton}) that
the value of the Clauser-Horne combination does not change when
all the four amplitudes are multiplied by the same phase factor.
We can therefore assume with no loss of generality that the
imaginary part of one of the amplitudes, which we will take to be
$\tilde{\alpha}_1$, is equal to zero. This leaves us with seven
independent real numbers that form the free working parameters for
the downhill simplex algorithm. As typical when optimizing
multidimensional non-linear functions that may possess multiple local
extrema, the downhill simplex algorithm does not guarantee to
provide the global extremum of the optimized function. In order to
improve the confidence of our results, we have therefore restarted
the algorithm several times with different initial conditions for
each optimization problem, and selected the best extremum. The
consistency of the obtained results suggests that the applied
procedure was adequate to the complexity of our problem.

In Fig.~\ref{foton-kombinacje} we depict the optimum violation
of the Clauser-Horne inequality as a function of the imperfection
parameters $\tilde{\eta}, \xi$, and $p_D$. As the dark count rates
of silicon avalanche photodiodes are relatively low, we have
chosen to study only two rather extreme
values of the zero dark count probability:
the ideal case $p_D=1$ and $p_D=0.99$ corresponding to $1\%$ dark
count rate, which is well above what can be expected from
modern detectors. It is seen that the difference between these
two cases is rather minor. Instead, we have focused our attention
on the detection efficiency $\tilde{\eta}$ and the mode mismatch $\xi$
as more critical parameters in a realistic situation. The graphs
shown in Fig.~\ref{foton-kombinacje} were obtained by selecting
a $50\times 50$ grid in the plotted range of $\tilde{\eta}$
and $\xi$, and by performing the optimization
procedure described above for each point of the grid.
The contours in the plots have been drawn
using interpolation, and the thick line separating the ``nonlocal''
region has been offset by $10^{-3}$ in order to avoid display of the effects
of numerical errors.

It is seen that the maximum value of the Clauser-Horne combination, obtained
for perfect detection efficiency and mode matching, lies above $0.15$.
As expected, this value diminishes when $\tilde{\eta}$
and/or $\xi$ decrease, thus implying more imperfections in the setup.
One can observe a clear trade-off between these two
quantities: the worse the detection efficiency is, the better
mode matching must be achieved in order to obtain a violation of
the Clauser-Horne inequality. The minimum detection efficiency
needed to observe the violation assuming perfect mode matching
and zero dark counts is about $84\%$.
In Fig.~\ref{foton-amplitudy} we show the values of the coherent
displacements that maximize the value of the Clauser-Horne combination.
Numerical calculations have shown that in most of the area the
maximum value is achieved by amplitudes that are real and pairwise equal:
$\tilde{\alpha}_1=\tilde{\beta}_1$ and $\tilde{\alpha}_2=\tilde{\beta}_2$.
These two values are depicted in Fig.~\ref{foton-amplitudy}(a).
Only in a small triangular
region in the parameter plane the situation becomes more complicated
and the Clauser-Horne combination is maximized by displacement parameters
with non-trivial imaginary parts. This region is shown in detail in
seven graphs in Fig.~\ref{foton-amplitudy}(b).

\section{Two-mode squeezed vacuum state}
\label{Sec:TwoModeSqueezedVacuum}

In this section we will discuss another possibility of using
spontaneous parametric down-conversion as a source of entanglement
for homodyne tests of Bell's inequalities. The single-photon state
discussed previously was obtained by conditional detection on
the down-conversion output in the weak conversion regime, when
the probability of generating simultaneously two or more photon
pairs can be safely neglected. In a general case, the two-mode
quantum state generated in spontaneous parametric down-conversion
is of the form:
\begin{equation}
\label{Nopa}
|\Phi\rangle = \frac{1}{\cosh r} \sum_{n=0}^\infty
\tanh^n r \,|n\rangle_A |n\rangle_B,
\end{equation}
where $r$ is a squeezing parameter including information about
the pump laser field, interaction time, thickness of the nonlinear
crystal, etc. We have assumed here that the twin beams are generated
in single spatio-temporal modes travelling towards directions $A$
and $B$.

The state given in Eq.~(\ref{Nopa}) clearly exhibits
entanglement in the Fock basis when the two modes are chosen
as the separate subsystems. It is therefore interesting to
study correlations between ``photon silence'' events
on two spatially separated unbalanced homodyne detectors fed
by the beams $A$ and $B$. We shall analyze whether such
correlations are capable of violating the Clauser-Horne
inequality (\ref{Eq:CH}). The relevant probabilities are
calculated in Appendix, with the final results given by:
\begin{widetext}
\begin{eqnarray}
\label{nopacorrel} Q(\tilde{\alpha}, \tilde{\beta} )&=&
\frac{p_{\text{D}}^2} {1+\tilde{\eta}(2-\tilde{\eta})\sinh^2 r}
\exp\left[
-\left(\frac{1+\tilde{\eta}\sinh^2 r}
{1+\tilde{\eta}(2-\tilde{\eta})\sinh^2 r}+
\frac{1-\xi}{\xi}\right)
\tilde{\eta}(|\tilde{\alpha}|^2 +
|\tilde{\beta}|^2 )
\right.
\nonumber \\
& &
\left.
+\frac{\tilde{\eta}^2\sinh r
\cosh r} {1+\tilde{\eta}(2-\tilde{\eta})\sinh^2 r}
(\tilde{\alpha}\tilde{\beta}+
\tilde{\alpha}^*\tilde{\beta}^*)\right]
\end{eqnarray}
\end{widetext}
for the probability of a joint ``photon silence'' event and
\begin{eqnarray}
Q(\tilde{\alpha}) & = & \
\frac{p_{\text{D}}}{1+\tilde{\eta}\sinh^2 r}
\nonumber \\
\label{Eq:nopasingle}
& & \times
\exp\left[ -\left(\frac{1} {1+\tilde{\eta}\sinh^2
r}+\frac{1-\xi}{\xi}\right)\tilde{\eta} |\tilde{\alpha}|^2
\right]
\end{eqnarray}
describing the marginal probability of a zero count event
on a single detector. Using these expressions we can build
the combination (\ref{Eq:CHCombination}) and check in what
regime the Clauser-Horne inequality can be violated.

In order to find the optimal violation of the Clauser-Horne
inequality we have followed the numerical procedure described in
Sec.~\ref{Sec:SinglePhotonState}. It is easy to note that the
Clauser-Horne combination is invariant with respect to multiplying
$\tilde{\alpha}_1, \tilde{\alpha}_2$ and $\tilde{\beta}_1,
\tilde{\beta}_2$ by conjugate phase factors; we can therefore
assume with no loss of generality that the amplitude
$\tilde{\alpha}_1$ is purely real. An additional parameter that we
have subjected to numerical optimization is the squeezing
parameter $r$, as it should be possible to tune it rather easily
by adjusting the intensity of the beam pumping the nonlinear
medium.

In Fig.~\ref{sciskanie-kombinacje} we depict the minimized value
of the Clauser-Horne combination as a function of the overall
efficiency $\tilde{\eta}$ and mode matching $\xi$, for two
values of the dark count rate: the ideal case $p_D=1$, and the 1\% dark
count probability corresponding to $p_D=0.99$.
Once again one can observe a trade-off
between the detection efficiency and the mode-matching. However,
it is interesting to note that when a two-mode squeezed
state is used as a source of entanglement, the minimum detection
efficiency needed to violate the Clauser-Horne inequality is
about $71\%$ in the case of perfect mode matching and absence
of dark counts. For completeness, we show also the optimized
values of the coherent displacements in Fig.~\ref{sciskanie-amplitudy}
and the squeezing parameter in Fig.~\ref{sciskanie-er}.
We have found numerically that the Clauser-Horne combination
is minimized by pairs of coherent displacements with opposite signs:
$\tilde{\alpha}_1 = - \tilde{\beta}_1$ and $\tilde{\alpha}_2
= - \tilde{\beta}_2$. It is also
interesting to note that optimal values of squeezing parameter
are rather moderate, which means that only the first several
terms in the Fock-basis expansion in Eq.~(\ref{Nopa}) are relevant.
This observation may be of experimental relevance, as good mode
matching between the signal and local oscillator fields may be
more difficult to achieve for strongly squeezed fields.

As it is well known, the two-mode squeezed vacuum state generated in spontaneous
parametric down-conversion is described by a positive definite Wigner function.
Quantum states of this form have been recalled on several occasions when discussing
to what extent the quantum mechanical phase space formalism can serve as a local
hidden-variable model. As this issue has generated some rather confusing comments
in previous works, it is instructive to analyze from this point of view our homodyne
scheme for testing Bell's inequalities.
For this purpose, let us use the Wigner formalism
to rewrite the probability of a joint
``photon silence'' event $Q(\tilde{\alpha}, \tilde{\beta})$ to a form
that resembles a local hidden variable model:
\begin{equation}
\label{Eq:QWigner}
Q(\tilde{\alpha}, \tilde{\beta})
=
\int \text{d}^2\lambda_1 \, \text{d}^2 \lambda_2 \;
\rho (\tilde{\alpha} ; \lambda_1 ) \rho (\tilde{\beta} ; \lambda_2 )
W ( \lambda_1, \lambda_2 )
\end{equation}
Here $\rho (\tilde{\alpha} ; \lambda_1 )$ and
$\rho (\tilde{\beta} ; \lambda_2 )$ are Wigner representations of the observables
corresponding to no-count events on the detectors $A$ and $B$, and
$W ( \lambda_1, \lambda_2 )$ is the positive definite
Wigner function of the state $|\Phi\rangle$. An easy calculation combining
Eq.~(\ref{jointsilence}) with Eq.~(\ref{familytransform}) from Appendix
shows that $\rho (\tilde{\alpha} ; \lambda_1 )$ is explicitly given by
\begin{equation}
\label{Eq:rhoalphalambda}
\rho (\tilde{\alpha} ; \lambda_1 ) = \frac{2 p_D}{2 - \tilde{\eta}}
\exp \left( - \frac{2\tilde{\eta}}{2-\tilde{\eta}} (\lambda_1
- \tilde{\alpha})^2 - \tilde{\eta} \frac{1-\xi}{\xi} | \tilde{\alpha} |^2
\right)
\end{equation}
with an analogous expression for $\rho (\tilde{\beta} ; \lambda_2
)$. In order to interpret $\rho (\tilde{\alpha} ; \lambda_1 )$ and
$\rho (\tilde{\beta} ; \lambda_2 )$ as local realities that are
defined by hidden variables $\lambda_1$ and $\lambda_2$, the
expression given in Eq.~(\ref{Eq:rhoalphalambda}) must be bounded
between $0$ and $1$ as representing a local probability of a
zero-count detection event. It is straightforward to see that the
upper bound is in general violated: in the perfect case of
$\tilde{\eta} = \xi = p_D = 1$ the maximum value attained by $\rho
(\tilde{\alpha} ; \lambda_1 )$ is equal to 2. Consequently, the
Wigner representation given in Eq.~(\ref{Eq:QWigner}) cannot be
usually interpreted as a local hidden variable model for the
``photon silence'' events. It is clearly seen from the above
example that positivity of the Wigner function describing a
composite quantum system is in principle no obstacle to use it for
demonstrating a violation of Bell's inequalities. An equally
important factor is whether the measured observables have a Wigner
representation that can be interpreted in local realistic terms.
Only when both these conditions are met, the detected correlation
functions have to satisfy Bell's inequalities.

\section{Conclusions}
\label{Sec:Conclusions}

In this paper, we presented a complete multimode theory
of measuring quantum optical quasidistribution functions by photon
counting. We demonstrated that by constructing a coherent field
whose excited modes match pairwise the signal modes of interest,
it is possible to scan point-by-point complete quasidistribution
functions. Our analysis included effects of experimental imperfections,
such as losses of the signal field, non-unit mode matching, and
accidental dark counts.

As an interesting application of the developed formalism, we performed
a feasibility study of using unbalanced homodyning for testing Bell's
inequalities. As a source of entanglement, we analyzed two states whose
generation based on the process of
spontaneous parametric down-conversion seems
presently to be most practical. We found the range of experimental
parameters that enable one to observe a violation of Bell's
inequalities without the standard postselection procedure. The minimum
detection efficiency required for such a loophole-free test, in the absence
of any other imperfections, is $84\%$ for a one-photon Fock state,
and $71\%$ for a two-mode squeezed vacuum state. Especially the latter
value seems to be within the reach of current technologies \cite{KwiaSteiPRA93}.
Our calculations demonstrated a clear trade-off between two parameters
describing experimental imperfections that are likely to determine
the performance of a realistic setup: the detection efficiency and the
mode mismatch. Such a trade-off is easily understandable: a natural way
to improve the modal structure of the signal field is to perform
spatio-temporal filtering, which, however, inevitably generates excess
losses of the signal field. This effect underlines the need to develop
sources of down-converted radiation with controlled spatio-temporal
structure \cite{BanaURenOpL01}.

Though it does not affect directly tests of Bell's inequalities,
it is worthwhile to note that the optical states used in the
scheme studied here do not share problems that are currently
inherent to the standard down-conversion sources of polarization
entangled states. In the latter case, the entangled photon pairs
are generated only with a small probability per the time slot
defined by the shape of the laser pulse pumping the nonlinear
medium. Thus, the actual state of the produced light is given,
up to the first order of the perturbation theory, by a small
two-photon contribution superposed with a strong vacuum
component \cite{KokBrauPRA00}.  Within current technology it is
not practical to perform a nondestructive test for the presence
of photons thus selecting {\em a priori} the photon-pair term.
Consequently experiments utilizing this type of entanglement,
especially in quantum information processing applications,
are usually based on the postselection of the detection events,
and the strong vacuum component to the produced states is
removed by retaining only the cases when a sufficient number of
photons has been registered by the detectors. One might hope
that this difficulty could be overcome by producing multiple
photon pairs and using the methods of entanglement swapping and
conditional detection to prepare the maximal polarization
entanglement in the event-ready manner \cite{ZukoZeilPRL93}.
However, there exists a rather general proof \cite{KokBrauPRA00}
suggesting that this route is highly difficult, if not
impossible, within presently available means. Practical
realization of the states used in our study is not affected by
these difficulties.  The single photon state divided on a 50:50
beam splitter can be produced by conditional detection performed
on the down-conversion output. Receiving signal from the trigger
detector tells us unambiguously that the required single-photon
state has been prepared with high fidelity. The preparation of
the second state that was considered here, i.e.\ the
two-mode squeezed vacuum state, does not require any
postselection at all. The state produced from vacuum in the
process of nondegenerate parametric down-conversion possess
photon-number entanglement which is sufficient to violate Bell's
inequalities. Of course, this feature is not critical in tests
of Bell's inequalities which can rely validly on random sources
of entangled photon pairs. However, the fact that we can reveal
quantum nonlocality by performing dichotomic measurements on
continuous-variable systems suggests that continous variable 
entanglement may be a useful resource in quantum information
processing protocols
based on binary logic, and our ability to produce it on demand
could be its important advantage.

\section*{\uppercase{Acknowledgements}}
We thank M. \.{Z}ukowski for valuable comments on the manuscript.
We would like to acknowledge useful discussions with I. A. Walmsley,
S. Wallentowitz, and W. Vogel. This research was supported by
ARO-administered MURI Grant No.\ DAAG-19-99-1-0125. A.D. thanks
European Science Foundation for a Short-Term Felowship under QIT
Programme. We also acknowledge the financial contribution of the
European Commission through the Research Training
Network QUEST and of the KBN Grant No.\ PBZ-KBN-043/P03/2001.

\appendix
\section*{Appendix}

Direct calculation of the quasidistributions of the state
(\ref{Nopa}) is not as straightforward as in the case of single
photon entangled with vacuum. Therefore we will calculate them
using another method. First we will recall a useful property of a
general, $M$-dimensional family of quasiprobability distributions:
from any $s$-ordered quasidistribution one can calculate any other
$s'$-ordered one using the following integral formula:
\begin{widetext}
\begin{equation}
\label{familytransform} W(\alpha_1,\ldots,\alpha_N;s') = \left(
\frac{2}{\pi(s-s')} \right)^M\, \int
\text{d}^2\beta_1\ldots\text{d}^2\beta_M\,
\exp\left(-\frac{2}{s-s'} \sum_{i=1}^M |\alpha_i-\beta_i|^2
\right) W(\beta_1,\ldots,\beta_M;s).
\end{equation}
A two-mode $Q$ function corresponding to the
quasidistribution ordering parameter $s=-1$ can be very
easily calculated for the NOPA state directly
from the Fock representation:
\begin{equation}
\label{Eq:WABgammadelta}
W_{AB}(\gamma,\delta;-1) = \frac{1}{\pi^2} \mid
\langle\gamma|\langle\delta|\Phi\rangle \mid^2 = \frac{1}{\pi^2
\cosh^2 r} \exp\left[-|\gamma|^2-|\delta|^2+ \tanh r\, \left(
\gamma\delta+\gamma^*\delta^* \right)\right].
\end{equation}
We will also need its marginal:
\begin{equation}
W_A(\gamma;-1) = \int \text{d}^2\delta\,
W_{AB} (\gamma,\delta;-1)
= \frac{1}{\pi\cosh^2 r}
\exp\left(-\frac{|\gamma|^2}{\cosh^2 r} \right).
\end{equation}
Inserting Eq.~(\ref{Eq:WABgammadelta}) into Eq.~(\ref{familytransform})
specialized to the two-mode case with $M=2$ one easily obtains that:
\begin{eqnarray}
W_{AB} \left( \tilde{\alpha} , \tilde{\beta} ;
- \frac{2- \tilde{\eta}}{\tilde{\eta}}\right) & = &
\frac{\tilde{\eta}^2}{\pi^2(1+\tilde{\eta}(2-\tilde{\eta})\sinh^2 r)}
\exp\left(
-\frac{\tilde{\eta}+\tilde{\eta}^2\sinh^2 r}
{1+\tilde{\eta}(2-\tilde{\eta})\sinh^2 r}
(|\tilde{\alpha}|^2+|\tilde{\beta}|^2)
\right. \nonumber \\
& &
\left.
+\frac{\tilde{\eta}^2\sinh r \cosh r}
{1+\tilde{\eta}(2-\tilde{\eta})\sinh^2 r}
(\tilde{\alpha}\tilde{\beta}+
\tilde{\alpha}^*\tilde{\beta}^*)\right)
\end{eqnarray}
\end{widetext}
Inserting the above formula into Eq.~(\ref{jointsilence}) yields
the probability of the joint ``photon silence'' event given
in Eq.~(\ref{nopacorrel}). An analogous calculation yields
the marginal single-mode quasidistribution function:
\begin{equation}
W_{A} \left(\tilde{\alpha};
-\frac{2-\tilde{\eta}}{\tilde{\eta}}\right)
= \frac{\tilde{\eta}}{\pi}
\frac{1}{1+\tilde{\eta}\sinh^2 r}
\exp\left( -\frac{\tilde{\eta}\,|\tilde{\alpha}|^2}
{1+\tilde{\eta}\sinh^2 r} \right).
\end{equation}
which after multiplying by the state-independent factors given in
Eq.~(\ref{singlesilence}) gives the marginal probability of a no-count event
given in Eq.~(\ref{Eq:nopasingle}).

\clearpage

\begin{widetext}

\begin{figure}
\begin{center}
\epsfig{file=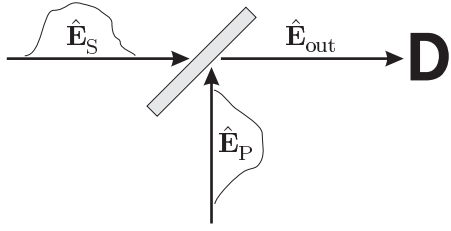}
\end{center}
\caption{\label{Fig:General} Experimental arrangement consisting of signal and
probe fields, interfered at a beam splitter characterized by a power
transmission $T$. The output port of the beam splitter is monitored by a photon
counting detector integrating the incident light over its active surface.}
\end{figure}

\begin{figure}
\begin{center}
\epsfig{file=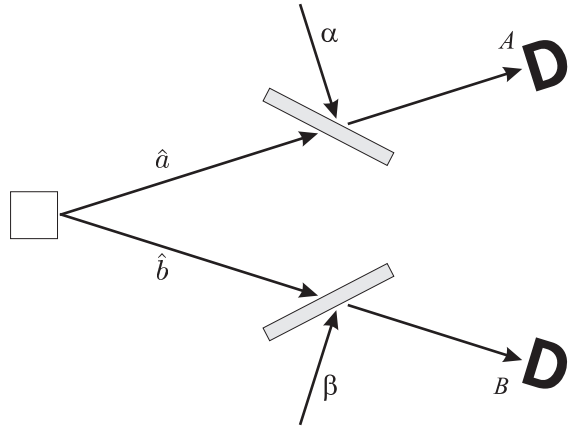}
\end{center}
\caption{\label{Fig:TwoMode} Two entangled and spatially separated light beams
fall onto unbalanced homodyne detectors composed of photon counters preceded by
high-transmission beam splitters with auxiliary coherent fields entering
through the sideway input ports.}
\end{figure}

\begin{figure}
\begin{center}
\epsfig{file=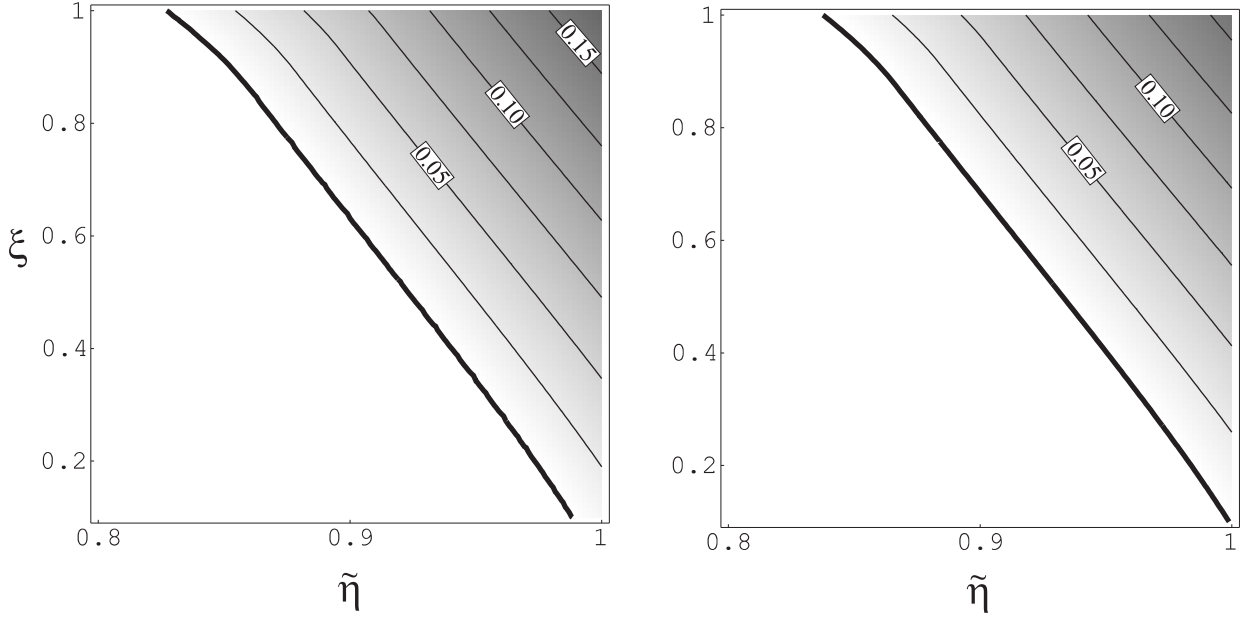}
\end{center}
\caption{\label{foton-kombinacje} Maximized Clauser-Horne
combination ${\cal CH}$ for a single photon divided at a beam splitter
as a function of the overall losses $\tilde{\eta}$ and the mode mismatch $\xi$. Plots
for noiseless detection $p_{\text{D}}=1$ (left) and with a
dark-count rate $p_{\text{D}}=0.99$ (right). The thick
lines represents the bound ${\cal CH}>0$ imposed by
local-realistic theories.}
\end{figure}

\begin{figure}
\begin{center}
\epsfig{file=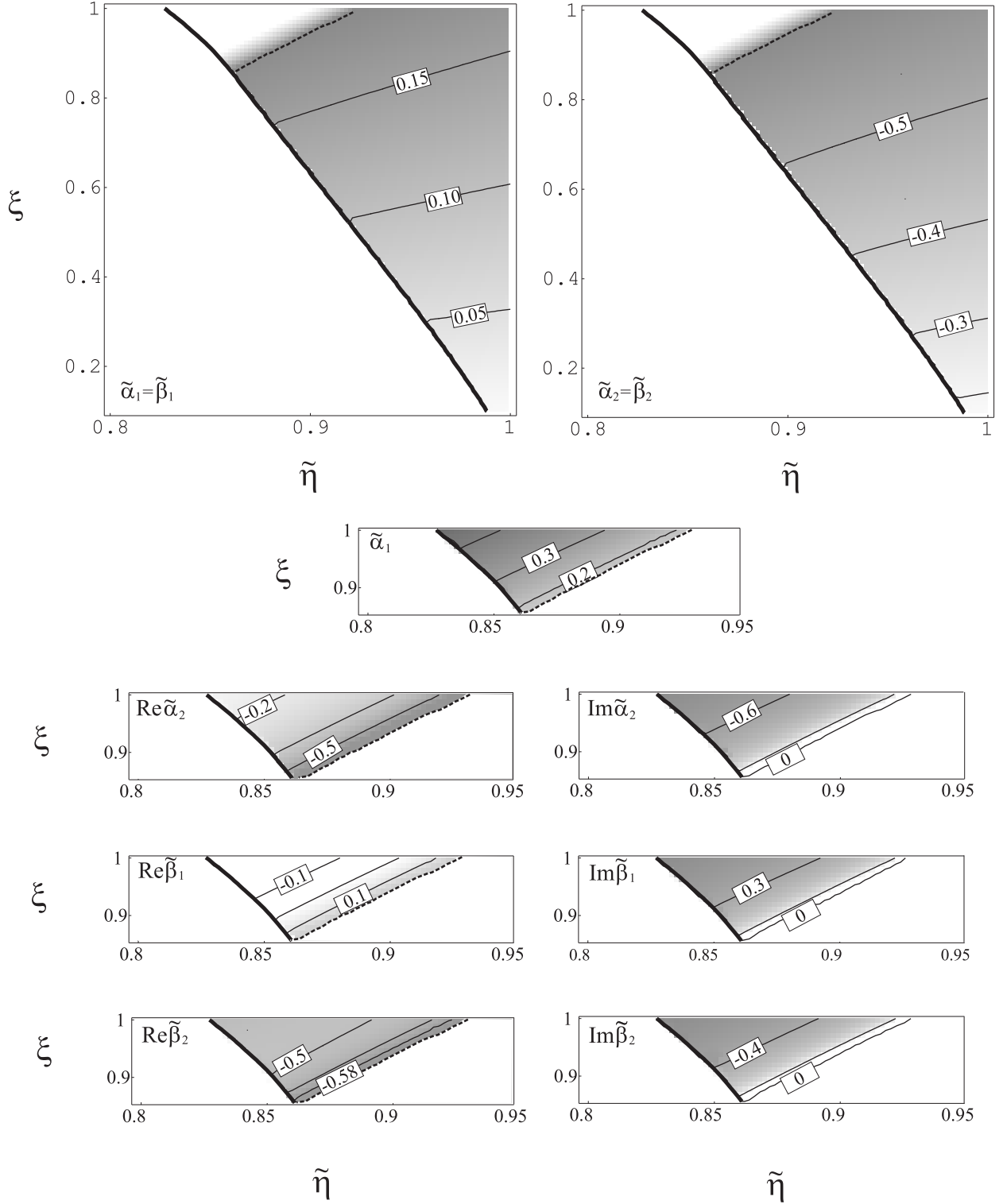}
\end{center}
\caption{\label{foton-amplitudy} Coherent amplitudes maximizing
violation of the Clauser-Horne inequality for the noiseless
detection ($p_{\text{D}}=1$) as a function
of the overall losses $\tilde{\eta}$ and the mode mismatch $\xi$.
For the most of the area of violation it is
sufficient to use just two real amplitudes (top
$\tilde{\alpha}_1=\tilde{\beta}_1$ and
$\tilde{\alpha}_2=\tilde{\beta}_2$. In a small
triangular area separated by a dashed line the maximum
violation is obtained for complex
amplitudes, shown in the bottom seven figures.}
\end{figure}

\begin{figure}
\begin{center}
\epsfig{file=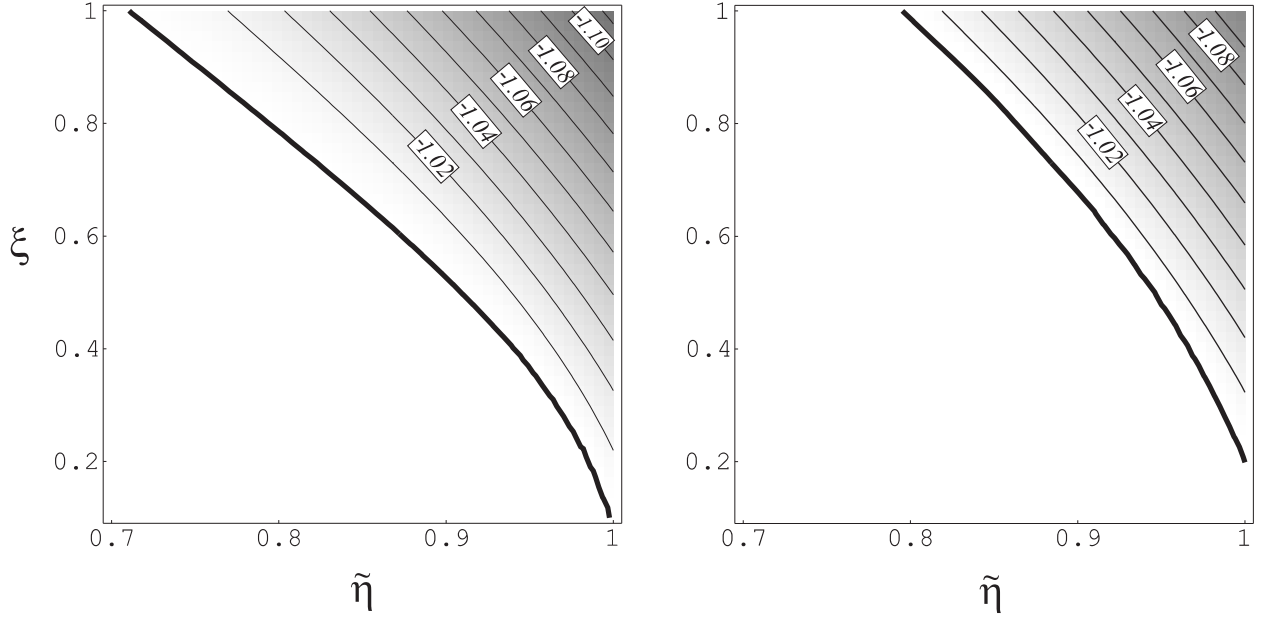}
\end{center}
\caption{\label{sciskanie-kombinacje} Optimal violation of the
Clauser-Horne inequality for a squeezed state in the case of
noiseless detection $p_{\text{D}}=1$ (left) and with a
dark-count rate $p_{\text{D}}=0.99$ (right), as a function
of the overall losses $\tilde{\eta}$ and the mode mismatch $\xi$.}
\end{figure}

\begin{figure}
\begin{center}
\epsfig{file=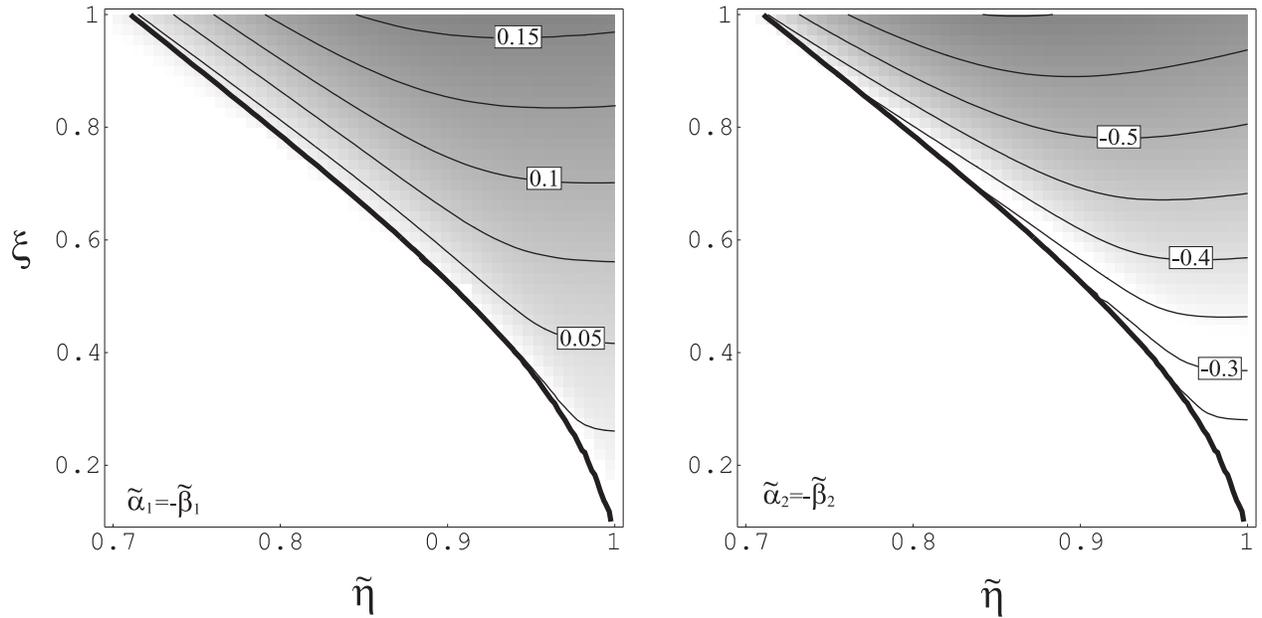}
\end{center}
\caption{\label{sciskanie-amplitudy} Coherent amplitudes
maximizing violation of the Clauser-Horne inequality for the
noiseless detection ($p_{\text{D}}=1$). It is sufficient to use just two real
amplitudes $\tilde{\alpha}_1=-\tilde{\beta}_1$ and
$\tilde{\alpha}_2=-\tilde{\beta}_2$.}
\end{figure}

\begin{figure}
\begin{center}
\epsfig{file=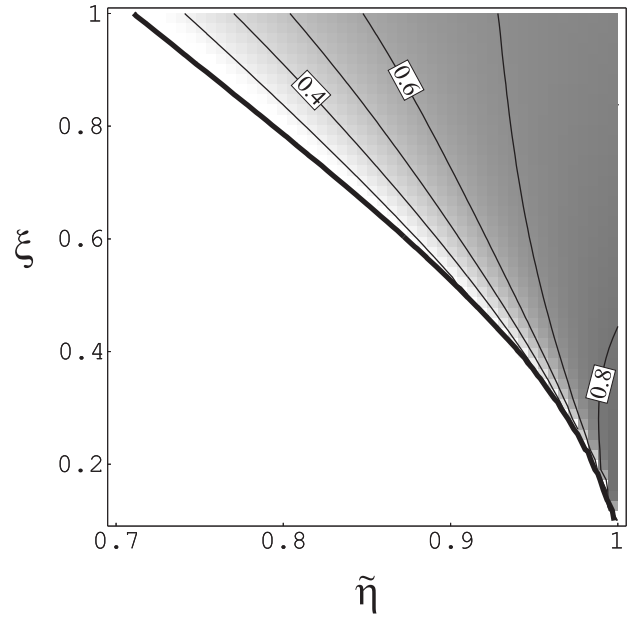}
\end{center}
\caption{\label{sciskanie-er} Squeezing parameter $r$
maximizing violation of the Clauser-Horne inequality for the
noiseless detection ($p_{\text{D}}=1$).}
\end{figure}

\clearpage
\end{widetext}

\end{document}